# Measurement of In-Band Optical Noise Spectral Density

Sylvain Almonacil, Matteo Lonardi, Philippe Jennevé *and* Nicolas Dubreuil

**We present a method to measure the spectral density of in-band optical transmission impairments without coherent electrical reception and digital signal processing at the receiver. We determine the method's accuracy by numerical simulations and show experimentally its feasibility, including the measure of in-band nonlinear distortions power densities.**

I. INTRODUCTION

UBIQUITUS and accurate measurement of the noise power, and its spectral characteristics, as well as the determination and quantification of the different noise sources are required to design future dynamic, low-margin, and intelligent optical networks, especially in open cable design, where the optical line must be intrinsically characterized.

In optical communications, performance is degraded by a plurality of impairments, such as the amplified spontaneous emission (ASE) due to Erbium doped-fiber amplifiers (EDFAs), the transmitter-receiver (TX-RX) imperfection noise, and the power-dependent Kerr-induced nonlinear impairments (NLI) [1]. Optical spectrum-based measurement techniques are routinely used to measure the out-of-band optical signal-to-noise ratio (OSNR) [2]. However, they fail in providing a correct assessment of the signal-to-noise ratio (SNR) and in-band noise statistical properties. Whereas the ASE noise is uniformly distributed in the whole EDFA spectral band, TX-RX noise and NLI mainly occur within the signal band [3]. Once the latter impairments dominate, optical spectrum-based OSNR monitoring fails to predict the system performance [4].

Lately, the scientific community has significantly worked on assessing the noise spectral characteristics and their impact on the SNR, trying to exploit the information in the digital domain by digital signal processing (DSP) or machine learning. In this direction, several efforts have been carried out to evaluate, measure, and model noise spectrum (or temporal correlation) induced by NLI [4-7]. Moreover, the discrimination between the various sources of impairments has been investigated, focusing on distinguishing linear ASE-related and NLI effects [7-12]. Although accurate measurements of in-band noise statistics are presently executed in the digital domain, they cannot provide the noise's complete spectral characterization in a distributed way. On the other hand, optical spectrum-based techniques are presently incapable of addressing in-band noise features such as NLI contributions, i.e., failing to infer performances accurately.

To the best of our knowledge, we describe hereafter the first performance measurement of in-band optical impairments spectral statistics, enabling to experimentally assess NLI spectral density characterization. The method is implemented in the optical frequency domain [2] by shaping the optical signal power spectral densities (PSDs) to cancel PSD contributions within a narrow spectral range, creating a controlled spectral hole in the signal spectrum. As it will be discussed hereafter, the hole width is selected to minimize the impact of the hole on the signal-dependent NLI. During propagation, the spectral hole is progressively filled by in-line noise and optical distortion contributions, for which the accumulated power is measured with an optical spectrum analyzer (OSA). By iteratively shifting the spectral hole position, the entire in-band optical noise PSD is reconstructed. This technique offers various beneficial traits. It operates under multi-impairments involved in the network, including various NLI contributions. As it does not require the full-coherent receiver, it is rate and modulation independent. Although we validate the technique on a single channel system as a proof-of-concept, it can be also employed in a multi-channel system. Moreover, as OSAs can be distributed ubiquitously along the optical path, this technique allows per-span and per-wavelength assessment. Finally, the capability of determining pure-optical impairments, i.e., of the optical line without transceiver DSP equalization and TR-RX noise, may facilitate the design and characterization of open submarine cables [13]. We could employ the technique to monitor performance. Yet, due to the degradation caused by spectral holes in live transmissions, the method is preferred for diagnostic or dimensioning purposes.

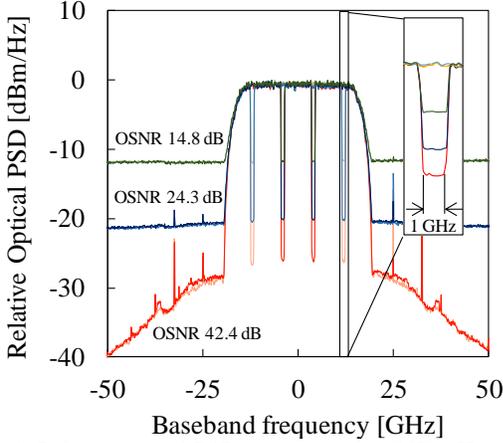

Figure 1: Relative optical PSDs (referenced to 0dBm/Hz for max spectral power) with different ASE noise levels for a 32.5 GBd signal with holes of width 1 GHz at $\pm 4$ GHZ and $\pm 12$ GHZ around the carrier frequency.

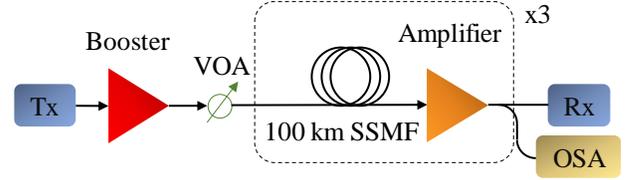

Figure 2: experimental set-up. VOA: variable optical attenuator, SSMF: standard single-mode fiber, OSA: optical spectrum analyzer. AWGN: additive white Gaussian noise.

The paper is organized as follows. In Section II, we detail the operating principle of the optical noise measurement technique. In Section III, we report on the measurements of in-band impairment spectral density in a 300 km long single-channel transmission, which exhibit a very good agreement with simulations and highlight the capability of the technique in assessing in-band colored NLI spectral contribution.

## II. ASSESSING IN-BAND NOISE WITH SPECTRAL HOLES

The in-band optical noise assessment with spectral holes aims to determine the PSD related to impairments affecting a given channel. It requires an optical TX and an OSA. The signal with a spectrum containing the hereafter so-called "spectral holes" is first generated. To do so, the fast Fourier transform (FFT) of each tributary of dual-polarization digital traces is computed. The FFT magnitude is set to 0 within the spectral range of two symmetric spectral holes, and an inverse FFT is performed to retrieve the new traces. In the case of non-symmetric spectral holes, one must operate in the complex digital field rather than in each tributary. This method relies on modern optical transmitters using high-speed digital-to-analog converters (DAC) and DSP that can generate custom spectral shapes. By iteratively changing the spectral holes' location within the signal spectrum and measuring the optical PSD value at the spectral holes' frequencies with an OSA, the complete noise optical PSD can be reconstructed without the need of a full coherent receiver.

Optical spectra for a 32.5 GBd signal with two symmetric 1 GHz wide holes located at $\pm 4$ GHZ and $\pm 12$ GHZ from the carrier frequency are shown in Fig. 1. The spectral hole width is set to 1 GHz to be larger than the OSA resolution, which is of the order of hundreds of MHz for typical high-resolution OSA based on coherent detection [14]. As we show later, the hole width is small enough to not alter the spectral contribution of the signal dependent NLI dramatically. The signal spectra have been generated with a block size that corresponds to the entire DAC memory length of $2^{18}$ samples, and a frequency resolution of the FFT/iFFT block equal to 460 kHz at 82 GSamples/s. At a symbol rate, $R$, and with root-raised-cosine signals of roll-off $\beta$, roughly $R(1 + \beta)/\Delta h$ different spectra must be generated and recorded to cover the whole signal spectrum. For instance, at 32.5 GBd with a roll-off factor of 0.2, it corresponds to about 40 different iterations. Nevertheless, we might divide by two this value by generating pairs of spectral holes centered around the carrier frequency. This symmetry also has the advantage of making the method immune to the transmitter in-phase/quadrature skew and imbalance, which would lead to modification of the PSD within non-symmetric spectral holes. It must be noted that the spectral holes generation is done prior to quantization, but given the 8-bit resolution of standard DACs, the quantization noise PSD is maintained well below the other sources of noise and does not impact the assessment of in-line distortions. In addition, the OSA instrumental noise contribution can be neglected considering the operating SNR values for standard optical transmission.

To illustrate the method, the measured spectra for a signal impaired by TX and ASE noises only are displayed in Fig.1 for 3 different OSNR values adjusted with an EDFA followed by a variable optical attenuator (VOA). As shown in the inset of Fig. 1, the accumulated noise is revealed by measuring the optical PSD within the spectral hole. Under a dominant ASE white noise impairment (OSNR 24.3 and 14.8 dB), the noise level measured in the spectral holes and outside the signal bandwidth is equal, confirming that the holes do not affect signal-independent noise spectral densities. It is worth noting that these measurements do not take the RX noise into account but solely the in-line distortions. However, TX-RX noise contribution can be straightforwardly determined through a transponder back-to-back profiling calibration.

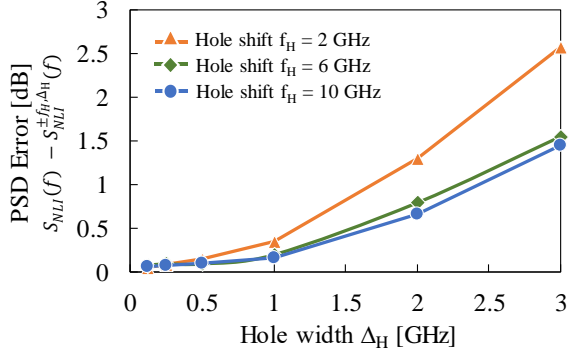

Figure 3 Variation of the PSD measurement errors with the spectral hole width for three different symmetric spectral hole positions respectively to the carrier frequency. Error is defined as the difference between the simulated PSD at the spectral hole position without and with a hole in the signal spectrum.

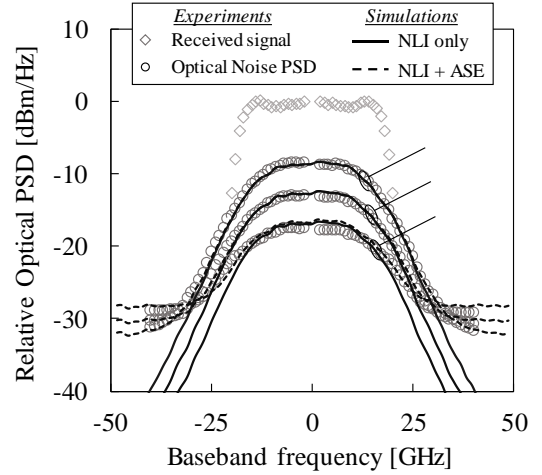

Figure 4: Experimental optical PSD of the RX signal (diamonds) and of the relative noise obtained with 1 GHz-wide spectral holes and for input power set to 9, 11 and 13 dBm (circles). Comparison with simulated relative optical PSDs in case of a pure NLI (solid lines) and a mixed NLI + ASE (dashed lines) impairments.

### III. PROOF OF CONCEPT

The capability of the method in accurately measuring and characterizing noise spectral density and transmission SNR is demonstrated for a point-to-point single-channel transmission by comparing measured noise PSDs with both DSP-based measurements at the receiver and simulations. For such transmission, it is customary to consider, as the three main sources of independent impairments [15], the ASE generated by EDFA, the Kerr-induced NLI, and the TX-RX imperfection noise.

*A. Experiment setup*

We consider the single channel point-to-point link depicted in Fig. 2. TX and the RX are connected through three 100 km-long standard single-mode fibers (SSMF) without inline chromatic dispersion compensation. We use a dual-polarization optical transmitter based on an 82 GS/s 8-bit DAC. The signal is a $2^{16}$ Debruijn sequence mapped on polarization-division multiplexed (PDM) quadrature phase-shift keying (QPSK) symbols and resampled at 2.5 samples per symbol, i.e., 32.5 GBd. The generated optical signal is boosted with an EDFA at 21 dBm and connected to an attenuator to adjust the input optical power launched into the fiber link between 5 and 13 dBm. At the RX, 1% of the optical power is sent to an OSA with 180 MHz resolution for noise optical PSD measurement and 99% to a wavelength selective switch to filter off the out-band noise. Finally, a coherent receiver with 40 GHz balanced photodiodes is used for demodulation and SNR measurements. The demodulation is performed in the same signal but without spectral holes. The photocurrents feed a 40 GS/s real-time oscilloscope with 20 GHz analog bandwidth. The DSP consists in resampling at two samples per symbol, compensating for chromatic dispersion (-5010 ps/nm), equalizing and demultiplexing polarizations with constant modulus algorithm, down sampling at 1 sample per symbol. It follows carrier frequency estimation and carrier phase recovery. For the phase recovery, we perform Viterbi-Viterbi algorithm with 301-tap average window to not compensate for nonlinear phase correlations [4]. After the symbol decision, the system performance is evaluated through the SNR computation based on a constellation analysis over $10^6$ symbols.

The optical PSD assessment is achieved by generating signal spectra with 1 GHz-wide spectral holes whose frequencies are scanned over a $\pm 20$ GHz spectral range respectively to the carrier frequency, with a 2 GHz increment. Every PSDs point is the average over five acquisitions.

*B. Simulation setup*

In order to compare with the experimental results, we simulated the propagation of a 32.5 GBd PDM-QPSK signal along the 300 km-long link with Optilux [16] over a simulation bandwidth of 100 GHz. Thanks to optical fiber characterization, the linear attenuation is set to 0.22 dB/km, and the dispersion coefficient to +16.70 ps/nm/km at 1550 nm with a slope of 0.057 ps/nm²/km. The nonlinear coefficient $\gamma$ has been taken equal to $1.22 \times 10^{-3}$ W$^{-1}$.m$^{-1}$ by fitting the measured performance vs. power curve at high power levels [17], matching with the reported values for standard SSMF. Simulations differ from experiments by the fact that in-line amplifiers are noiseless and fully recover fiber attenuation. As customary in simulations, we account or not for linear noise by lumped noise loading according to the out-band noise level collected during experiments, i.e., an additive white Gaussian noise loading with $P_{\text{LIN}}^{\text{sim}} \simeq$ -19 dBm flat over all the simulation bandwidth. Concerning the TX-RX noise contribution, we added to the performance curves a constant $SNR_{\text{TX-RX}}^{-1}$ = -19.15 dB, obtained by transceiver profiling calibration similarly to [15]. Moreover, dispersion is perfectly recovered by an attenuation-less fully linear fiber. Received signal and noise spectra are estimated with the periodogram method using $2^8$ points rectangular window and normalized to the spectrum power levels observed at the OSA. For the sake of clarity, we report the results in the NLI dominated regime only, i.e., where optical spectrum-based methods fail in predicting actual SNR and noise spectral characteristics.

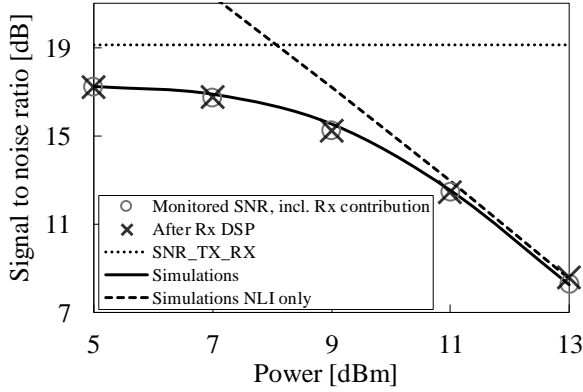

Figure 5 Signal-to-noise ratio obtained by spectral hole-based method (circles) accounting for the RX noise, constellation analysis of the received samples at the decision gate (crosses), and simulations (lines).

*C. Impact of the Spectral Holes*

To quantify the impact of the hole-based method on the nonlinear spectrum's characterization, we simulated NLI PSDs for a signal with and without spectral holes, defined respectively as $S_{NLI}^{\pm f_H, \Delta_H}(f)$ and $S_{NLI}(f)$, with $f$ the frequency, $\pm f_H$ the symmetrical hole center frequencies around the carrier and $\Delta_H$ the hole width. We compare simulations for various hole widths and shifts within the signal bandwidth, and we used a larger number of periodogram points for a better resolution ($2^{13}$). Figure 3 shows PSD errors sampled at the hole shift center, $\epsilon^{\pm f_H, \Delta_H}(f_H) = S_{NLI}(f_H) - S_{NLI}^{\pm f_H, \Delta_H}(f_H)$, in terms of spectral hole widths for three different $f_H = \pm 2, \pm 6$ and $\pm 10$ GHz. As expected, the hole broadening leads to a higher PSD error. For instance, regarding a $\pm 2$ GHz-shifted holes, a 3 GHz-wide hole impacts the measurement with an error estimated equal to 2.5 dB while it is reduced below 0.5 dB for widths smaller than 1 GHz. Moreover, for a fixed width $\Delta_H$, the perturbation of the noise PSD is larger for spectral holes positioned closer to the carrier frequency. Note that the method always underestimates noise PSD and that both the holes' positions and widths influence the measurement's accuracy. Although the insertion of the spectral hole modifies the NLI that take place in the fiber, this self-action can be constrained by carefully setting the spectral hole width in regards of the needed accuracy and OSA resolution.

*D. Results and observations*

Based on the previous simulations, and to reach a trade-off between PSD characterization and performance assessment, we set the hole width to 1 GHz in the experiments. The noise optical PSDs measured for input powers equal to 9, 11, and 13 dBm, i.e., operating in a regime where the NLI is dominant within the signal bandwidth, are plotted in Fig. 4 with circles. The in-band noise PSD contributions are revealed by measuring the optical PSD value at the spectral holes' frequencies with the OSA. The relative PSDs are given respectively to the received optical spectrum measured for a signal without spectral holes and plotted with diamonds. For the sake of clarity, the received spectrum in Fig. 4 is truncated within the signal bandwidth ($\pm 20$ GHz) since the out-band contributions for the reported input powers are superimposed with their respected optical noise PSD displayed with circles.

The measured optical noise PSDs achieved throughout the proposed method are compared with simulated noise PSDs calculated without holes in the signal spectrum. In Fig. 4, the pure nonlinear and mixed linear-nonlinear PSDs simulated for 9, 11, and 13 dBm launch powers are reported respectively with solid and dashed lines. The empirically measured optical noise PSDs show good agreement with the simulated in-band and out-band noise contributions. This confirms that the 1 GHz spectral hole width does not influence the NLI spectrum significantly. We can observe an explicit nonlinear behavior concerning the in-band noise contributions since NLI dominates all other noise sources. Additionally, a 2 dB difference in the launch power leads to a 4 dB variation in the measured optical noise PSDs, as reported by the simulated spectra, according to the expected nonlinear behaviors.

The previous optical PSD noise measurements enable to determine the SNR, accounting for the additional RX noise contribution derived by a back-to-back calibration analysis. Figure 5 shows the SNR variations with the launch power obtained with the spectral hole based-method corrected by the additional RX-noise contribution (circles). For comparison, crosses indicate SNR measured at the decision gate by constellation analysis, while the solid line represents simulations. The excellent matching between these curves demonstrates the proposed spectral hole-based method's ability to accurately assess and estimate in-band colored NLI spectral contribution. Furthermore, to highlight that the performance curves account for all transmission impairments (ASE, NLI, and TX), the two asymptotic curves for NLI (slope 2 dB/dB) and TX-RX noise (slope 0 dB/dB) are plotted in Fig. 5 with dashed lines.

## IV. CONCLUSIONS

We propose, for the first time, a method able of appraising the in-band optical distortion spectral characteristics of a fiber transmission channel. The procedure relies on modifying the TX PSD by inserting narrow spectral holes. Thanks to a proof of concept, we show excellent agreement between experiments and simulations. With this method, we can observe the in-band optical noise PSD in the presence of dominant and colored NLI, which is impracticable with traditional optical spectrum-based noise methods. Finally, we also prove the successful all-optical assessment of the SNR in the NLI governed regime.